\documentclass[reprint, amsmath,amssymb, showpacs, aps]{revtex4-1}

\usepackage{graphicx}
\usepackage{dcolumn}
\usepackage{bm}
       
\begin{document}

\title{Magneto-optical imaging technique for hostile environment: the ghost imaging approach.}

\author{A.~Meda} \affiliation{INRIM, Strada delle Cacce 91, I-10135 Torino, Italy}

\author{A.~Caprile} \affiliation{INRIM, Strada delle Cacce 91, I-10135 Torino, Italy}

\author{A.~Avella} \affiliation{INRIM, Strada delle Cacce 91, I-10135 Torino, Italy}

\author{I.~Ruo ~Berchera} \affiliation{INRIM, Strada delle Cacce 91, I-10135 Torino, Italy}

\author{I.~P.~Degiovanni} \affiliation{INRIM, Strada delle Cacce 91, I-10135 Torino, Italy}

\author{A.~Magni} \affiliation{INRIM, Strada delle Cacce 91, I-10135 Torino, Italy}

\author{M.~Genovese} \affiliation{INRIM, Strada delle Cacce 91, I-10135 Torino, Italy}

\begin{abstract}
We develop a new approach in magneto-optical imaging (MOI), applying for the first time a ghost imaging (GI) protocol to perform Faraday microscopy.
MOI is of the utmost importance for the investigation of magnetic properties of material samples, through Weiss domains shape, dimension and dynamics analysis.
Nevertheless, in some extreme conditions such as e. g. cryogenic temperatures or high magnetic fields application, there exists a lack of domains images due to the difficulty in creating an efficient imaging system in such environments. Here we present an innovative MOI technique that separates the imaging optical path from the one illuminating the object. The technique is based on thermal light GI and exploits correlations between light beams to retrieve the image of magnetic domains. As a proof of principle, the proposed technique is applied to the Faraday magneto-optical observation of the remanence domain structure of an yttrium iron garnet sample.
\end{abstract}

\maketitle
\emph{Introduction.} The magnetic behavior of a material is directly correlated to its domain structure (DS). The DS is determined by, amongst other things, chemical composition, microstructure, growth processes and post-process treatments. The capability to observe the DS is fundamental for designing tuned material properties and to obtain the desired magnetic behavior \cite{man10}.
The understanding of the structure and dynamics of magnetic domains and domain walls is of the utmost importance in a number of applications, such as thin-film recording heads in the magnetic recording industry \cite{all05} and spin-electronic devices in information technology \cite{cha07}. Data about magnetic anisotropy, internal stresses, magnetic losses and domain wall dynamics can be inferred by magnetic structure imaging \cite{mag11,mag12}.
Several imaging techniques have been developed over the years, e.g. magneto-optical Kerr and Faraday microscopy, magnetic force microscopy (MFM), magnetization sensitive electron microscopy, scanning Hall probe microscopy, scanning SQUID microscopy, and Bitter decoration.
Magneto-optical techniques \cite{sch98,arg01}, exploiting the Kerr effects in reflection and the Faraday effect in transmission, in particular, have some advantages, among which the ease of setup and a very good time resolution, allowing the investigation of fast dynamical processes for the characterization of magnets \cite{mag11} and superconductors \cite{joh04}.
Due to the Faraday effect, the polarization plane of the light transmitted through a magnetized material endowed with spontaneous local magnetization is rotated of an angle \cite{zve97}
$
\phi_{\pm} \approx \pm L \frac{\omega M_{\boldsymbol{k}}}{2c \sqrt{\varepsilon}},
$
where $L$ is the thickness of the magnetic material, $\omega$ is the light frequency, $c$ is the speed of light, $M_{\boldsymbol{k}}$ is the local magnetization component along the wavevector $\boldsymbol{k}$ and $\varepsilon$ is the dielectric permittivity of the medium. The sign of the angle $\phi$ depends on the sign of the magnetization.
Similarly, due to the magneto-optical Kerr effects, a rotation occurs when the light is reflected from the surface of a magnetized sample \cite{sch98}.
In a Faraday (Kerr) microscope, a beam of polarized light is exploited to perform the imaging of adjacent domains, with different magetization orientation, sending the transmitted (reflected) light to a
spatially resolving detector (e. g. a CCD) preceded by a polarization analyzer.
Such setup guarantees a magneto-optical contrast between adjacent domains.
The DS, observed by means of the magneto-optical effect, can be considered as an object with transmission $T(M_k(\mathbf{x}))$ (reflection $R$) depending on the local magnetization at the coordinate $\mathbf{x}$.
In the last years, with the advent of magnetocalorics earlier and spincaloritronics then \cite{bau12}, the possibility to perform magneto optical imaging in temperature variable experiments is acquiring in perspective more and more interest.
This means, unavoidably, to look for the possibility to integrate a magnetic imaging system in experimental setups where the optical access to the sample is strongly limited, due to temperature control additional stages \cite{kue15} or when the sample would be cooled at extremely low temperatures.
Until now, MOI at temperature down to 1K has been achieved by means of cryostats endowed with windows, which allow polarized light to reach the sample. However, radiation losses due to the presence of windows hinder the efficiency of the thermal isolation \cite{vil08}, limiting the performance of the cryostat in sub-Kelvin regime.
Furthermore, errors can be introduced due to the Faraday rotation in the microscope objective, with visible effects already at applied fields of the order of hundreds of mT, which unavoidably worsen when approaching the regime of several Teslas \cite{kuc14}.
Similarly, the applied magnetic field intensity could affect the correct operation of the camera electronics.
In this paper, we propose an innovative, flexible way of performing MOI of DS
exploiting the ghost imaging (GI) technique \cite{deg07, str95, bri11,val05,fer05,che10,sim14}. 
Compared to conventional techniques, in a GI setup the imaging system is placed in a spatially separated optical path with respect to the one containing the sample under test; the light-beam that interacts with the sample is collected by means of a bucket detector (a detector without spatial resolution) like, e.g., an optical fiber connected to a photodiode. A second beam, locally correlated with the probe beam, does not interact with the sample and is sent to a spatial resolving detector (CCD camera).
The technique retrieve the image of the DS exploiting correlations \cite{gat99,tan08,med13} between the output of the bucket detector with each pixel of the CCD.
This setup paves the way to realize MOI in extremely high magnetic fields, in window-less cryostats or whenever the accessible volume in the proximity of the sample is limited.

\emph{Theory of thermal Ghost Imaging.} Thermal GI (Figure \ref{GIscheme}) requires a couple of spatially incoherent, locally correlated thermal beams \cite{gla63,man95}, dubbed here as $a$-beam  and $b$-beam, generally obtained by splitting a single pseudo-thermal beam through a beam splitter.
In the two respective transverse planes, ideally divided in discrete resolution cells (each one larger than a "spatial coherence area") of coordinate $\mathbf{x}_{i}$  ($i=a,b$), the two beams present thermal intensity fluctuation statistically independent cell-by-cell,
\begin{equation}\label{variance n}
\left\langle (\delta n_{\mathbf{x}_{i}})^{2}\right\rangle= \left\langle n_{\mathbf{x}_{i}}\right\rangle\left( 1+\frac{\left\langle n_{\mathbf{x}_{i}}\right\rangle}{M}\right)\approx\frac{\left\langle n_{\mathbf{x}_{i}}\right\rangle^{2}}{M},
\end{equation}
where $\left\langle n_{\mathbf{x}_{i}}\right\rangle$ and $M$ represent the mean number of photons and the number of independent spatio-temporal modes collected within a cell during the acquisition time, respectively.
$\delta n_{\mathbf{x}_{i}} =n_{\mathbf{x}_{i}}-\langle n_{\mathbf{x}_{i}}\rangle$ is the photon number fluctuation. The last approximation in Eq. (\ref{variance n}), valid when $\left\langle n_{\mathbf{x}_{i}}\right\rangle / M \gg1$, is fulfilled in typical experimental configurations.
The mean photon number is conveniently expressed as
\begin{equation}\label{mean n}
\left\langle n_{\mathbf{x}_{i}}\right\rangle=T_{i}(\mathbf{x}_{i}) \lambda \quad i=a,b
\end{equation}
where $\lambda$ is the mean number of photons impinging on the resolution cell while $T_{i}(\mathbf{x}_{i})$ represents the overall locally defined transmission/detection efficiency of the channels.
In GI $a$-beam first interacts with the object with a structured transmission profile $\sim T_{a}(\mathbf{x}_{a})$, then is detected as a whole by the bucket detector, i.e. a detector without spatial resolution, whose output signal is $N_{a}=\sum_{\mathbf{x}_{a}} n_{\mathbf{x}_{a}}$ (the sum extends over a defined number of resolution cells).
The $b$-beam is detected by a spatially resolving detector at the correlated transverse plane, like a 2D-pixels array. The output signal from each pixel in the position $\mathbf{x}_{b}$ is denoted by $n_{\mathbf{x}_{b}}$.
The point-by-point spatial correlation, that emerges from the splitting of the intensity fluctuation, can be expressed in terms of the covariance of the
detected photon number as:
\begin{figure}[tbp]
\begin{center}
\includegraphics[angle=0, width=6cm]{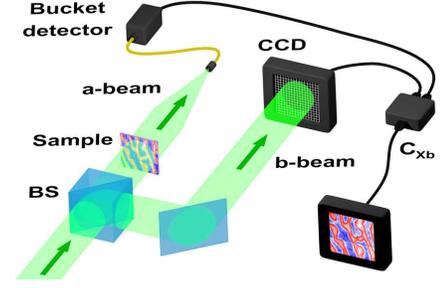}
\caption{Thermal Ghost imaging schematic representation: two correlated beams, $a$-beam and $b$-beam, are generated by sending an incoherent beam through a $50 \%$ beam splitter (BS). Then, beams are sent to two distinct optical path, one containing the sample to be imaged and a bucket detector, the other one containing a spatial resolving detector. The image of the sample is retrieved correlating the output of the two detectors. }
\label{GIscheme}
\end{center}
\end{figure}
\begin{equation}\label{cov nanb}
\left\langle \delta n_{\mathbf{x'}_{a}} \delta n_{\mathbf{x}_{b}}\right\rangle=\frac{\left\langle n_{\mathbf{x'}_{a}}\right\rangle \left\langle n_{\mathbf{x}_{b}}\right\rangle}{M}\delta_{\mathbf{x'}_{a},\mathbf{x}_{b}}= T_{b} T_{a}(\mathbf{x'}_{a})\frac{\lambda^2}{M} \delta_{\mathbf{x'}_{a}\mathbf{x}_{b}}
\end{equation}
where we have considered a uniform transmission-detection efficiency in the channel $b$, i.e. $T_{b}(\mathbf{x}_{b})=T_{b}$.
Therefore, the covariance of a single pixel in $\mathbf{x}_{b}$ with the signal from the bucket detector in channel $a$ turns out to be
\begin{equation}\label{cov Nanb}
\left\langle \delta N_{a} \delta n_{\mathbf{x}_{b}}\right\rangle=\sum_{\mathbf{x'}_{a}} \left\langle\delta n_{\mathbf{x'}_{a}}\delta n_{\mathbf{x}_{b}}\right\rangle = T_{b} T_{a}(\mathbf{x}_{b}) \frac{\lambda^2}{M},
\end{equation}
which permits to reconstruct on channel $b$ the transmission profile of the object interacting with beam $a$.
This is the essence of the GI technique.
Eq.s  (\ref{cov nanb}) and (\ref{cov Nanb}) show how GI reconstruction emerges point-by-point as a correlation of pairs of resolution cells immersed in a large number of uncorrelated contributions.
Thus, the statistical ensemble (namely the number of acquired photons $n_{\mathbf{x}_{b}}$ and of values $N_{a}$) should be "large enough" to allow the emergence of such correlation from the uncorrelated contribution; this increase the total duration of the experiment. Obviously, it is of utmost practical importance to quantify what does "large enough" mean for a specific GI application.
As usual, the theoretical mean values are estimated by an arithmetic average of $K$ samples from the same population, i.e. $\langle X\rangle\rightarrow E[X]=1/K \sum^{K}_{k=1}X_{k}$. In particular, the covariance in Eq. (\ref{cov Nanb}) is estimated by the statistical quantity
$C_{\mathbf{x}_{b}}=E[(N_a - E[N_a])(n_{\mathbf{x}_{b}} - E[n_{\mathbf{x}_{b}}])].$
As expected, its value converges to the theoretical one,
\begin{equation}\label{C Nanb}
\langle C_{\mathbf{x}_{b}}\rangle\simeq\langle\delta N_{a} \delta n_{\mathbf{x}_{b}}\rangle
\end{equation}
with fluctuations around the mean value of ($K>>1$)
\begin{eqnarray}\label{delta C Nanb}
\langle (\delta C_{\mathbf{x}_{b}})^{2}\rangle &\simeq & K^{-1}\langle (\delta N_{a} \delta n_{\mathbf{x}_{b}})^{2}\rangle-\langle \delta N_{a} \delta n_{\mathbf{x}_{b}}\rangle^{2}\\\nonumber
&\approx& K^{-1}\langle (\delta N_{a})^{2}\rangle \langle(\delta n_{\mathbf{x}_{b}})^{2}\rangle\\\nonumber
\end{eqnarray}
The approximation in the second line of Eq. (\ref{delta C Nanb}) holds since almost all the light collected by the bucket detector is uncorrelated with the pixel of the spatially resolving detector.
As expected, the fluctuation scales towards zero when increasing the sample size $K$.
Let us consider now an object defined by two regions $S_{a,j}$ ($j=+,-$) characterized by two different transmittance $T_{a,j}$ for $\mathbf{x}_{a}\in S_{a,j}$. On channel $b$ we can identify two regions $S_{b,j}$, each one locally correlated with the corresponding $S_{a,j}$.
The signal-to-noise ratio (SNR) (sometimes referred as to "contrast-to-noise ratio" in GI literature \cite{erk09,aga09, bas07,bri11}) defined as
\begin{equation} \label{SNRdef}
SNR = \frac{|\langle C_{+}- C_{-}\rangle|}{\sqrt{\langle (\delta C_ {+})^2\rangle+\langle (\delta C_{-})^2}}
\end{equation}
should be larger than 1.
In fact, by using Eq. (\ref{C Nanb}) with the substitution of Eq. (\ref{cov Nanb}) one gets $\langle C_{j}\rangle= T_{b} T_{a,j} \lambda^{2}/M $.
In the same way, the result in Eq. (\ref{delta C Nanb}) with the substitution of Eq.s (\ref{variance n}) and (\ref{mean n}) returns the evaluation of each noise component at the denominator of the SNR as $\langle (\delta C_ {j})^2\rangle\approx K^{-1} T^{2}_{a,j} R_{j} T_{b} \frac{\lambda^4}{M^2}$, leading to:
\begin{equation} \label{SNRevaluated}
SNR\approx \sqrt{K} \frac{|T_{a,+}- T_{a,-}|}{\sqrt{T^2_ {a,+} R_{+} + T^2_ {a,-} R_{-}}}.
\end{equation}
SNR only depends on the transmittance of the two regions $T_{a,j}$, on the number of resolution cells $R_{j}$ respectively transmitted through the object and on the size of the statistical sample $K$.
In particular, in the magneto-optical application presented in this work, we have roughly $R_{+}=R_{-}=R$. In this case it is easy to show that the SNR is maximized where one of the transmittance is null, reaching the optimal value $SNR = \sqrt{K/R}$.

\begin{figure}[tbp]
\begin{center}
\includegraphics[angle=0, width=8cm]{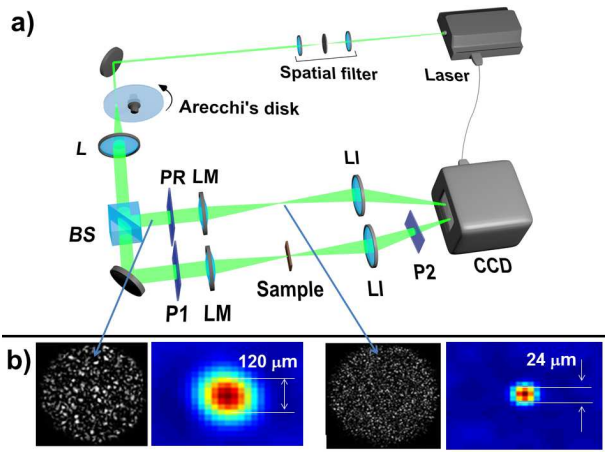}
\caption{(a) The magneto-optical GI experimental setup for Faraday microscopy of magnetic domains. In the scheme BS $=$ 50-50 beam splitter, PR. P1 and P2 are polarizers, L, LM and LI are lenses of focal length $f=100$ mm, $f_M=100$ mm and $f_I=75$ mm respectively. (b) The auto-correlation measurement for evaluating the coherence area, before and after the microscope with magnification factor $\mathbb{M}=0.2$.}
\label{setup}
\end{center}
\end{figure}

\emph{Faraday ghost imaging microscopy: experimental setup.}
The experimental setup of our magneto-optical GI experiment is depicted in Figure \ref{setup} (a).
The two locally correlated thermal beams, $a$-beam  and $b$-beam, are obtained by means of a single pseudo-thermal beam divided by a 50-50 beam splitter.
Polarized pseudo-thermal statistics light is engineered by sending coherent light pulses, such as a polarized pulsed laser beam, on a scattering medium with random scattering centers distribution, in order to obtain a speckle pattern with (pseudo) thermal intensity fluctuations.
In our case the coherent source is the second harmonic of a Q-switched Nd-YAG laser ($532$ nm), with pulse duration of $10$ ns and repetition rate of $12.4$ Hz. A spatial filter removes non-Gaussian components of the laser. Pulses are sent to a rotating ground-glass disk, known as Arecchi's disk \cite{are65} (with grain dimensions of some $\mu m$s).
The rotation speed is set in order to be considered frozen within the single pulse duration, but so that different pulses experience different independent grain distribution. Thus, each pulse generates an independent intensity spatial distribution, uncorrelated with the previous one.
A far-field speckle pattern is obtained with lens $L$, with $f=100$ mm, placed in $f-f$ configuration with respect to the disk.
A non-polarizing beam splitter separates the pseudo-thermal beam in two balanced, polarized and correlated pseudo-thermal beams, with spatial correlation of $98\%$ \cite{bri09}. The $a$-beam  is the transmitted beam while the $b$-beam is the reflected one.
$a$-beam is further polarized by means of the polarizer $P1$ and is sent to the sample, a film of YIG of chemical composition (YSmCa)$_3$(FeGe)$_5$O$_{12}$, characterized by out-of-plane anisotropy and an average domain size of 160 $\mu$m \cite{mag00}. The polarizer $P2$ allows magneto-optical contrast observation between adjacent domains.
In our sample, which is a simple case of uniaxial system at the remanence, the magnetization points along the direction perpendicular to the sample plane.
Adjacent domains have opposite magnetization, hence transmittance $T$, after proper polarization selection, can be considered as $T(\phi_+)=T_+$ or $T(\phi_-)=T_-$ in the case of positive or negative Faraday rotation angle respectively.
In our setup, both $a$ and $b$ beams are sent, by means of imaging lenses $LI$ of $f_I=75$ mm, to different portion of a CCD, with frame acquisition synchronized with laser pulses.
Bucket detector condition is fulfilled by summing CCD pixels in the region illuminated by $a$-beam.
The camera is an EMCCD (Andor Luca R camera), with high sensitivity and active area of $8 \times 8$ mm$^2$ and squared pixels of size $8 \mu$m.
For each laser shot, the output of the system is the number of photons $N_a$ detected by the bucket and the matrix of detected photons $n_{\mathbf{x}_{b}}$ corresponding to the pixel counts in the region of the CCD illuminated by $b$-beam.
The speckle dimension corresponds to the resolution cell dimension, and hence, to the resolution of the reconstructed image.
The diameter of the speckles is evaluated as the full width at half maximum (FWHM) of the auto-correlation coefficient estimated as \cite{bri09A}:
\begin{equation}
c(\mathbf{\xi})=\sum_{\mathbf{x}_{b}} \frac{ \delta n_{\mathbf{x}_{b}} \delta n_{\mathbf{x}_{b}+\mathbf{\xi}} }{\sqrt{[\delta n_{\mathbf{x}_{b}}]^2   [\delta n_{\mathbf{x}_{b}+\mathbf{\xi}}]^2}}   \label{A(xi)}
\end{equation}
where $\mathbf{\xi}$ is the two-dimensional shift of the selected region with respect to itself and $\mathbf{x}_{b}$ is the vector position of the pixel of the region $b$ (see Figure \ref{setup} (b)).
In order to enhance the resolution, the speckle dimension was reduced by increasing the diameter of the collimated coherent source impinging on the Arecchi's disk \cite{kla68} (see Figure \ref{setup} (b)). Also, a microscope with lenses $LM$ ($f_{M} = 100$ nm) able to further reduce the speckle diameter with a magnification factor $\mathbb{M}=0.2$ was constructed. In this way the number of speckles on the sample was increased. The setup finally provides a spatial resolution of $24$ $\mu$m at the plane of the object, largely sufficient to observe the magnetic domains of our sample.


\emph{Ghost imaging of magnetic domains.}
The ghost image is reconstructed evaluating the experimental covariance $C(\mathbf{x}_{b})$ averaging on the total number of frames $K$ acquired by the CCD.
We set polarizer $P2$ angle, minimizing transmittance $T_{-}$. In this way we are able to get as close as possible to the ideal condition for the SNR, as discussed after Eq. (\ref{SNRevaluated}).
We finally set $T_b$ with a polarizer $PR$ in order to avoid saturation of the CCD sensor; anyway, this has no effect on the image quality (see Eq. (\ref{SNRevaluated})).
\begin{figure}[tbp]
\begin{center}
\includegraphics[angle=0, width=8cm]{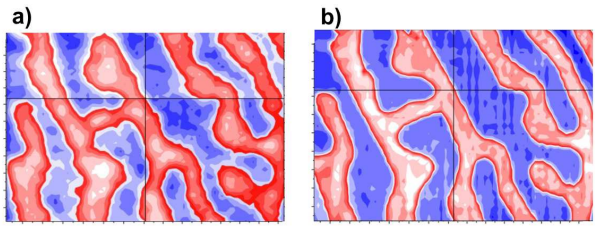}
\caption{(a) Ghost imaging of magnetic domains of a portion (a $(400$ x $560) \mu m^2$ area) of our YIG sample in comparison to (b) the direct Faraday imaging of the same portion }
\label{Result}
\end{center}
\end{figure}
In Figure \ref{Result} we compare GI of the Weiss domains with traditional Faraday imaging (TFI); for the chosen angle of the polarizer $P2$ we observe the intensity distribution $n_{\mathbf{x}_{a}}$ on the CCD area illuminated by $a$-beam, before summing pixels values for the bucket detector.
The images of the magnetic domains of our sample obtained with GI technique (a) and with TFI (b), averaging over $K =46904$ frames, include a $(400$ x $560) \mu m^2$ area characterized by equally distributed transmittances, such that $R_+$ = $R_-$ = $R$.
Each pixel of the ghost image corresponds to the value of the correlation $C_{\mathbf{x}_{b}}$ between $N_a$ and the $i$-th pixel measurement $n_{\mathbf{x}_{b}}$.
Weiss domains shape and position are evident in GI reconstruction, showing a good agreement with TFI. Differences between the two images are mostly related to a reduction of the resolution in GI. In TFI, the resolution of the image corresponds to the physical pixel dimension ($8$ $\mu$m) while for GI we evaluated $FWHM_{c(\mathbf{\xi})} \simeq 3$ pixels.
\begin{figure}[tbp]
\begin{center}
\includegraphics[angle=0, width=7cm]{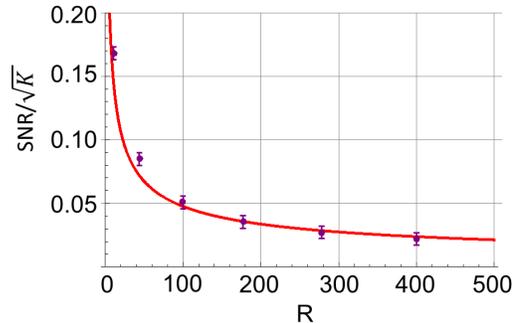}
\caption{SNR curve as a function of the number of resolution cell collected in the area with transmittance $T_{a,+}$. Red line correspond to the theoretical prediction.}
\label{SNR}
\end{center}
\end{figure}
Figure \ref{SNR} shows the SNR of the ghost image as a function of the number of speckles  $R$ collected in areas with transmittance $T_{a,+}$, showing a clear agreement with the theoretical prediction.
Red line refers to a theoretical model of SNR similar to Eq. \ref{SNRevaluated} but considering also a background contribution. In fact, in our case, since we estimate $T_{a,+} = 0.028 $, $T_{a,-} = 0.011 $ and  our pseudo-thermal light with $M=1$ provides a number of photons of $\lambda = 1210$, the detected mean intensity are quite low: $\overline{n}_{\mathbf{x}_{a}} = 34$ for $\mathbf{x}_{a}\in S_{a,+}$ and $\overline{n}_{\mathbf{x}_{a}} = 13$ for $\mathbf{x}_{a}\in S_{a,-}$. Therefore, background variance contribution ($V_{back} = 151$) is not negligible in GI reconstruction.

\emph{Conclusion.}
In this paper we proposed the application of the GI technique to magneto-optical Faraday (or Kerr) microscopy.
GI is a well established imaging technique in the realm of quantum optics, that allows retrieving the image of an object from an optical beam it has never interacted with. This is achieved by exploiting the correlation with its conjugated beam interacting with the object and observed with a bucket detector.
Due to its flexibility, this technique appears particularly promising if magnetic domains imaging has to be performed in hostile environment (very high magnetic field, sub-K regime, limited optical access to the sample), where the possibility of collecting the light after the interaction with the sample in a minimal amount of space, even without spatial resolution, e. g. by means of a single optical fiber, allows to overcome the limitations of the experimental setup.
To prove the validity of this technique in the field of MOI we performed a first proof-of-principle experiment exploiting the Faraday effect for imaging the domains of a YIG sample.
The techniques is extremely flexible and both Kerr or Faraday configurations can be exploited.
We achieved a resolution of $24$ $\mu$m by means of a purpose developed microscope but, with a proper design, the resolution of the ghost image can be increased at the level of traditional imaging.
Our results demonstrate that this technique is able to provide high quality images of magnetic domains.
In future, a system devoted to GI of magnetic domains to be used in hostile environment will be developed.
This will be based on the configuration known as "computational-ghost-imaging" \cite{sha08,bro09}, using a fibre-bundle to bring the multi-speckle-like light on the object, and a pigtailed photodiode with a multimode fiber as the bucket detector.

\textbf{Acknowledgments.}
This work has received funding from EU-FP7 BRISQ
project, JRP EXL02 - SIQUTE project (on the basis of
Decision No. 912/2009/EC), from MIUR (FIRB "LiCHIS" - RBFR10YQ3H and PRIN "DyNanoMag") and Progetto
Premiale "Oltre i limiti classici di misura"). We thank Dr. Marco Co\"{i}sson (INRIM) for useful discussions.

\bibliography{bibx}

\begin{thebibliography}{35}%
\makeatletter
\providecommand \@ifxundefined [1]{%
 \@ifx{#1\undefined}
}%
\providecommand \@ifnum [1]{%
 \ifnum #1\expandafter \@firstoftwo
 \else \expandafter \@secondoftwo
 \fi
}%
\providecommand \@ifx [1]{%
 \ifx #1\expandafter \@firstoftwo
 \else \expandafter \@secondoftwo
 \fi
}%
\providecommand \natexlab [1]{#1}%
\providecommand \enquote  [1]{``#1''}%
\providecommand \bibnamefont  [1]{#1}%
\providecommand \bibfnamefont [1]{#1}%
\providecommand \citenamefont [1]{#1}%
\providecommand \href@noop [0]{\@secondoftwo}%
\providecommand \href [0]{\begingroup \@sanitize@url \@href}%
\providecommand \@href[1]{\@@startlink{#1}\@@href}%
\providecommand \@@href[1]{\endgroup#1\@@endlink}%
\providecommand \@sanitize@url [0]{\catcode `\\12\catcode `\$12\catcode
  `\&12\catcode `\#12\catcode `\^12\catcode `\_12\catcode `\%12\relax}%
\providecommand \@@startlink[1]{}%
\providecommand \@@endlink[0]{}%
\providecommand \url  [0]{\begingroup\@sanitize@url \@url }%
\providecommand \@url [1]{\endgroup\@href {#1}{\urlprefix }}%
\providecommand \urlprefix  [0]{URL }%
\providecommand \Eprint [0]{\href }%
\providecommand \doibase [0]{http://dx.doi.org/}%
\providecommand \selectlanguage [0]{\@gobble}%
\providecommand \bibinfo  [0]{\@secondoftwo}%
\providecommand \bibfield  [0]{\@secondoftwo}%
\providecommand \translation [1]{[#1]}%
\providecommand \BibitemOpen [0]{}%
\providecommand \bibitemStop [0]{}%
\providecommand \bibitemNoStop [0]{.\EOS\space}%
\providecommand \EOS [0]{\spacefactor3000\relax}%
\providecommand \BibitemShut  [1]{\csname bibitem#1\endcsname}%
\let\auto@bib@innerbib\@empty
\bibitem [{\citenamefont {Manke}\ \emph {et~al.}(2010)\citenamefont {Manke},
  \citenamefont {Kardjilov}, \citenamefont {Schaefer}, \citenamefont {Hilger},
  \citenamefont {Strobl}, \citenamefont {Dawson}, \citenamefont {Gruenzweig},
  \citenamefont {G.Behr}, \citenamefont {Hentschel}, \citenamefont {David},
  \citenamefont {Kupsch}, \citenamefont {Lange},\ and\ \citenamefont
  {Banhart}}]{man10}%
  \BibitemOpen
  \bibfield  {author} {\bibinfo {author} {\bibfnamefont {I.}~\bibnamefont
  {Manke}}, \bibinfo {author} {\bibfnamefont {N.}~\bibnamefont {Kardjilov}},
  \bibinfo {author} {\bibfnamefont {R.}~\bibnamefont {Schaefer}}, \bibinfo
  {author} {\bibfnamefont {A.}~\bibnamefont {Hilger}}, \bibinfo {author}
  {\bibfnamefont {M.}~\bibnamefont {Strobl}}, \bibinfo {author} {\bibfnamefont
  {M.}~\bibnamefont {Dawson}}, \bibinfo {author} {\bibfnamefont
  {C.}~\bibnamefont {Gruenzweig}}, \bibinfo {author} {\bibnamefont {G.Behr}},
  \bibinfo {author} {\bibfnamefont {M.}~\bibnamefont {Hentschel}}, \bibinfo
  {author} {\bibfnamefont {C.}~\bibnamefont {David}}, \bibinfo {author}
  {\bibfnamefont {A.}~\bibnamefont {Kupsch}}, \bibinfo {author} {\bibfnamefont
  {A.}~\bibnamefont {Lange}}, \ and\ \bibinfo {author} {\bibfnamefont
  {J.}~\bibnamefont {Banhart}},\ }\href@noop {} {\bibfield  {journal} {\bibinfo
   {journal} {Nature communications}\ }\textbf {\bibinfo {volume} {1}}
  (\bibinfo {year} {2010})}\BibitemShut {NoStop}%
\bibitem [{\citenamefont {Allwood}\ \emph {et~al.}(2005)\citenamefont
  {Allwood}, \citenamefont {Xiong}, \citenamefont {Faulkner}, \citenamefont
  {Atkinson}, \citenamefont {Petit},\ and\ \citenamefont {Cowburn}}]{all05}%
  \BibitemOpen
  \bibfield  {author} {\bibinfo {author} {\bibfnamefont {D.~A.}\ \bibnamefont
  {Allwood}}, \bibinfo {author} {\bibfnamefont {G.}~\bibnamefont {Xiong}},
  \bibinfo {author} {\bibfnamefont {C.~C.}\ \bibnamefont {Faulkner}}, \bibinfo
  {author} {\bibfnamefont {D.}~\bibnamefont {Atkinson}}, \bibinfo {author}
  {\bibfnamefont {D.}~\bibnamefont {Petit}}, \ and\ \bibinfo {author}
  {\bibfnamefont {R.~P.}\ \bibnamefont {Cowburn}},\ }\href@noop {} {\bibfield
  {journal} {\bibinfo  {journal} {Science}\ }\textbf {\bibinfo {volume}
  {309}},\ \bibinfo {pages} {1688} (\bibinfo {year} {2005})}\BibitemShut
  {NoStop}%
\bibitem [{\citenamefont {Chappert}\ \emph {et~al.}(2007)\citenamefont
  {Chappert}, \citenamefont {Fert},\ and\ \citenamefont {Dau}}]{cha07}%
  \BibitemOpen
  \bibfield  {author} {\bibinfo {author} {\bibfnamefont {C.}~\bibnamefont
  {Chappert}}, \bibinfo {author} {\bibfnamefont {A.}~\bibnamefont {Fert}}, \
  and\ \bibinfo {author} {\bibfnamefont {F.~N.~V.}\ \bibnamefont {Dau}},\
  }\href@noop {} {\bibfield  {journal} {\bibinfo  {journal} {Nature Materials}\
  }\textbf {\bibinfo {volume} {6}},\ \bibinfo {pages} {813} (\bibinfo {year}
  {2007})}\BibitemShut {NoStop}%
\bibitem [{\citenamefont {Magni}\ \emph {et~al.}(2011)\citenamefont {Magni},
  \citenamefont {Fiorillo}, \citenamefont {Caprile}, \citenamefont {Ferrara},\
  and\ \citenamefont {Martino}}]{mag11}%
  \BibitemOpen
  \bibfield  {author} {\bibinfo {author} {\bibfnamefont {A.}~\bibnamefont
  {Magni}}, \bibinfo {author} {\bibfnamefont {F.}~\bibnamefont {Fiorillo}},
  \bibinfo {author} {\bibfnamefont {A.}~\bibnamefont {Caprile}}, \bibinfo
  {author} {\bibfnamefont {E.}~\bibnamefont {Ferrara}}, \ and\ \bibinfo
  {author} {\bibfnamefont {L.}~\bibnamefont {Martino}},\ }\href@noop {}
  {\bibfield  {journal} {\bibinfo  {journal} {Journ. of Appl. Phys.}\ }\textbf
  {\bibinfo {volume} {109}},\ \bibinfo {pages} {07A322} (\bibinfo {year}
  {2011})}\BibitemShut {NoStop}%
\bibitem [{\citenamefont {Magni}\ \emph {et~al.}(2012)\citenamefont {Magni},
  \citenamefont {Fiorillo}, \citenamefont {Ferrara}, \citenamefont {Caprile},
  \citenamefont {Bottauscio},\ and\ \citenamefont {Beatrice}}]{mag12}%
  \BibitemOpen
  \bibfield  {author} {\bibinfo {author} {\bibfnamefont {A.}~\bibnamefont
  {Magni}}, \bibinfo {author} {\bibfnamefont {F.}~\bibnamefont {Fiorillo}},
  \bibinfo {author} {\bibfnamefont {E.}~\bibnamefont {Ferrara}}, \bibinfo
  {author} {\bibfnamefont {A.}~\bibnamefont {Caprile}}, \bibinfo {author}
  {\bibfnamefont {O.}~\bibnamefont {Bottauscio}}, \ and\ \bibinfo {author}
  {\bibfnamefont {C.}~\bibnamefont {Beatrice}},\ }\href@noop {} {\bibfield
  {journal} {\bibinfo  {journal} {Magnetics, IEEE Transactions on}\ }\textbf
  {\bibinfo {volume} {48}},\ \bibinfo {pages} {3796} (\bibinfo {year}
  {2012})}\BibitemShut {NoStop}%
\bibitem [{\citenamefont {Hubert}\ and\ \citenamefont
  {Schaefer}(1998)}]{sch98}%
  \BibitemOpen
  \bibfield  {author} {\bibinfo {author} {\bibfnamefont {A.}~\bibnamefont
  {Hubert}}\ and\ \bibinfo {author} {\bibfnamefont {R.}~\bibnamefont
  {Schaefer}},\ }\href@noop {} {\emph {\bibinfo {title} {Magnetic Domains}}}\
  (\bibinfo  {publisher} {Berlin: Springer},\ \bibinfo {year}
  {1998})\BibitemShut {NoStop}%
\bibitem [{\citenamefont {Argyle}\ and\ \citenamefont {Mccord}(2001)}]{arg01}%
  \BibitemOpen
  \bibfield  {author} {\bibinfo {author} {\bibfnamefont {B.~E.}\ \bibnamefont
  {Argyle}}\ and\ \bibinfo {author} {\bibfnamefont {J.~G.}\ \bibnamefont
  {Mccord}},\ }\href {\doibase 10.1007/978-94-010-0624-8_22} {\emph {\bibinfo
  {title} {Magnetic Storage Systems Beyond 2000}}},\ edited by\ \bibinfo
  {editor} {\bibfnamefont {G.}~\bibnamefont {Hadjipanayis}},\ \bibinfo {series}
  {NATO Science Series}, Vol.~\bibinfo {volume} {41}\ (\bibinfo  {publisher}
  {Springer Netherlands},\ \bibinfo {year} {2001})\ pp.\ \bibinfo {pages}
  {287--305}\BibitemShut {NoStop}%
\bibitem [{\citenamefont {Johansen}\ and\ \citenamefont
  {Shantsev}(2004)}]{joh04}%
  \BibitemOpen
  \bibfield  {author} {\bibinfo {author} {\bibfnamefont {T.~H.}\ \bibnamefont
  {Johansen}}\ and\ \bibinfo {author} {\bibfnamefont {D.~V.}\ \bibnamefont
  {Shantsev}},\ }\href@noop {} {\emph {\bibinfo {title} {Magneto-optical
  imaging}}}\ (\bibinfo  {publisher} {Springer Science \& Business Media},\
  \bibinfo {year} {2004})\BibitemShut {NoStop}%
\bibitem [{\citenamefont {Zvezdin}\ and\ \citenamefont {Kotov}(1997)}]{zve97}%
  \BibitemOpen
  \bibfield  {author} {\bibinfo {author} {\bibfnamefont {A.~K.}\ \bibnamefont
  {Zvezdin}}\ and\ \bibinfo {author} {\bibfnamefont {V.~A.}\ \bibnamefont
  {Kotov}},\ }\href@noop {} {\emph {\bibinfo {title} {Modern Magnetooptics and
  Magnetooptical Materials}}}\ (\bibinfo  {publisher} {IOP British and
  Philadelphia},\ \bibinfo {year} {1997})\BibitemShut {NoStop}%
\bibitem [{\citenamefont {Bauer}\ and\ \citenamefont {van Wees}(2012)}]{bau12}%
  \BibitemOpen
  \bibfield  {author} {\bibinfo {author} {\bibfnamefont {G.~E.~W.}\
  \bibnamefont {Bauer}}\ and\ \bibinfo {author} {\bibfnamefont {E.~S. B.~J.}\
  \bibnamefont {van Wees}},\ }\href {\doibase 10.1038/nmat3301} {\bibfield
  {journal} {\bibinfo  {journal} {Nature Materials}\ }\textbf {\bibinfo
  {volume} {11}},\ \bibinfo {pages} {391} (\bibinfo {year} {2012})}\BibitemShut
  {NoStop}%
\bibitem [{\citenamefont {Sola}\ \emph {et~al.}(2015)\citenamefont {Sola},
  \citenamefont {Kuepferling}, \citenamefont {Basso}, \citenamefont {Pasquale},
  \citenamefont {Kikkawa}, \citenamefont {Uchida},\ and\ \citenamefont
  {Saitoh}}]{kue15}%
  \BibitemOpen
  \bibfield  {author} {\bibinfo {author} {\bibfnamefont {A.}~\bibnamefont
  {Sola}}, \bibinfo {author} {\bibfnamefont {M.}~\bibnamefont {Kuepferling}},
  \bibinfo {author} {\bibfnamefont {V.}~\bibnamefont {Basso}}, \bibinfo
  {author} {\bibfnamefont {M.}~\bibnamefont {Pasquale}}, \bibinfo {author}
  {\bibfnamefont {T.}~\bibnamefont {Kikkawa}}, \bibinfo {author} {\bibfnamefont
  {K.}~\bibnamefont {Uchida}}, \ and\ \bibinfo {author} {\bibfnamefont
  {E.}~\bibnamefont {Saitoh}},\ }\href@noop {} {\bibfield  {journal} {\bibinfo
  {journal} {Jour. of Appl. Phys.}\ }\textbf {\bibinfo {volume} {117}},\
  \bibinfo {pages} {17C510} (\bibinfo {year} {2015})}\BibitemShut {NoStop}%
\bibitem [{\citenamefont {Villaume}\ \emph {et~al.}(2008)\citenamefont
  {Villaume}, \citenamefont {Antonevici}, \citenamefont {Bourgault},
  \citenamefont {Leggeri}, \citenamefont {Porcar},\ and\ \citenamefont
  {Villard}}]{vil08}%
  \BibitemOpen
  \bibfield  {author} {\bibinfo {author} {\bibfnamefont {A.}~\bibnamefont
  {Villaume}}, \bibinfo {author} {\bibfnamefont {A.}~\bibnamefont
  {Antonevici}}, \bibinfo {author} {\bibfnamefont {D.}~\bibnamefont
  {Bourgault}}, \bibinfo {author} {\bibfnamefont {J.~P.}\ \bibnamefont
  {Leggeri}}, \bibinfo {author} {\bibfnamefont {L.}~\bibnamefont {Porcar}}, \
  and\ \bibinfo {author} {\bibfnamefont {C.}~\bibnamefont {Villard}},\
  }\href@noop {} {\bibfield  {journal} {\bibinfo  {journal} {Rev. Sci.
  Instrum.}\ }\textbf {\bibinfo {volume} {79}},\ \bibinfo {pages} {023904}
  (\bibinfo {year} {2008})}\BibitemShut {NoStop}%
\bibitem [{\citenamefont {Kuch}\ \emph {et~al.}(2015)\citenamefont {Kuch},
  \citenamefont {Schaefer}, \citenamefont {Fischer},\ and\ \citenamefont
  {Hillebrecht}}]{kuc14}%
  \BibitemOpen
  \bibfield  {author} {\bibinfo {author} {\bibfnamefont {W.}~\bibnamefont
  {Kuch}}, \bibinfo {author} {\bibfnamefont {R.}~\bibnamefont {Schaefer}},
  \bibinfo {author} {\bibfnamefont {P.}~\bibnamefont {Fischer}}, \ and\
  \bibinfo {author} {\bibfnamefont {F.~U.}\ \bibnamefont {Hillebrecht}},\
  }\href@noop {} {\emph {\bibinfo {title} {Magnetic Microscopy of Layered
  Structures}}}\ (\bibinfo  {publisher} {Springer Series in Surface Sciences},\
  \bibinfo {year} {2015})\BibitemShut {NoStop}%
\bibitem [{\citenamefont {Degiovanni}\ \emph {et~al.}(2007)\citenamefont
  {Degiovanni}, \citenamefont {Bondani}, \citenamefont {Puddu}, \citenamefont
  {Andreoni},\ and\ \citenamefont {Paris}}]{deg07}%
  \BibitemOpen
  \bibfield  {author} {\bibinfo {author} {\bibfnamefont {I.}~\bibnamefont
  {Degiovanni}}, \bibinfo {author} {\bibfnamefont {M.}~\bibnamefont {Bondani}},
  \bibinfo {author} {\bibfnamefont {E.}~\bibnamefont {Puddu}}, \bibinfo
  {author} {\bibfnamefont {A.}~\bibnamefont {Andreoni}}, \ and\ \bibinfo
  {author} {\bibfnamefont {M.~G.~A.}\ \bibnamefont {Paris}},\ }\href@noop {}
  {\bibfield  {journal} {\bibinfo  {journal} {Phys. Rev. A}\ }\textbf {\bibinfo
  {volume} {76}},\ \bibinfo {pages} {062309} (\bibinfo {year}
  {2007})}\BibitemShut {NoStop}%
\bibitem [{\citenamefont {Strekalov}\ \emph {et~al.}(1995)\citenamefont
  {Strekalov}, \citenamefont {Sergienko}, \citenamefont {Klyshko},\ and\
  \citenamefont {Shih}}]{str95}%
  \BibitemOpen
  \bibfield  {author} {\bibinfo {author} {\bibfnamefont {D.~V.}\ \bibnamefont
  {Strekalov}}, \bibinfo {author} {\bibfnamefont {A.~V.}\ \bibnamefont
  {Sergienko}}, \bibinfo {author} {\bibfnamefont {D.~N.}\ \bibnamefont
  {Klyshko}}, \ and\ \bibinfo {author} {\bibfnamefont {Y.~H.}\ \bibnamefont
  {Shih}},\ }\href {http://link.aps.org/doi/10.1103/PhysRevLett.74.3600}
  {\bibfield  {journal} {\bibinfo  {journal} {Phys. Rev. Lett.}\ }\textbf
  {\bibinfo {volume} {74}},\ \bibinfo {pages} {3600} (\bibinfo {year}
  {1995})}\BibitemShut {NoStop}%
\bibitem [{\citenamefont {Brida}\ \emph {et~al.}(2011)\citenamefont {Brida},
  \citenamefont {Chekhova}, \citenamefont {Fornaro}, \citenamefont {Genovese},
  \citenamefont {Lopaeva},\ and\ \citenamefont {Berchera}}]{bri11}%
  \BibitemOpen
  \bibfield  {author} {\bibinfo {author} {\bibfnamefont {G.}~\bibnamefont
  {Brida}}, \bibinfo {author} {\bibfnamefont {M.~V.}\ \bibnamefont {Chekhova}},
  \bibinfo {author} {\bibfnamefont {G.~A.}\ \bibnamefont {Fornaro}}, \bibinfo
  {author} {\bibfnamefont {M.}~\bibnamefont {Genovese}}, \bibinfo {author}
  {\bibfnamefont {E.~D.}\ \bibnamefont {Lopaeva}}, \ and\ \bibinfo {author}
  {\bibfnamefont {I.~R.}\ \bibnamefont {Berchera}},\ }\href {\doibase
  10.1103/PhysRevA.83.063807} {\bibfield  {journal} {\bibinfo  {journal} {Phys.
  Rev. A}\ }\textbf {\bibinfo {volume} {83}},\ \bibinfo {pages} {063807}
  (\bibinfo {year} {2011})}\BibitemShut {NoStop}%
\bibitem [{\citenamefont {Valencia}\ \emph {et~al.}(2005)\citenamefont
  {Valencia}, \citenamefont {Scarcelli}, \citenamefont {D�Angelo},\ and\
  \citenamefont {Shih}}]{val05}%
  \BibitemOpen
  \bibfield  {author} {\bibinfo {author} {\bibfnamefont {A.}~\bibnamefont
  {Valencia}}, \bibinfo {author} {\bibfnamefont {G.}~\bibnamefont {Scarcelli}},
  \bibinfo {author} {\bibfnamefont {M.}~\bibnamefont {D�Angelo}}, \ and\
  \bibinfo {author} {\bibfnamefont {Y.}~\bibnamefont {Shih}},\ }\href@noop {}
  {\bibfield  {journal} {\bibinfo  {journal} {Phys. Rev. Lett.}\ }\textbf
  {\bibinfo {volume} {94}} (\bibinfo {year} {2005})}\BibitemShut {NoStop}%
\bibitem [{\citenamefont {Ferri}\ \emph {et~al.}(2005)\citenamefont {Ferri},
  \citenamefont {Magatti}, \citenamefont {Gatti}, \citenamefont {Bache},
  \citenamefont {Brambilla},\ and\ \citenamefont {Lugiato}}]{fer05}%
  \BibitemOpen
  \bibfield  {author} {\bibinfo {author} {\bibfnamefont {F.}~\bibnamefont
  {Ferri}}, \bibinfo {author} {\bibfnamefont {D.}~\bibnamefont {Magatti}},
  \bibinfo {author} {\bibfnamefont {A.}~\bibnamefont {Gatti}}, \bibinfo
  {author} {\bibfnamefont {M.}~\bibnamefont {Bache}}, \bibinfo {author}
  {\bibfnamefont {E.}~\bibnamefont {Brambilla}}, \ and\ \bibinfo {author}
  {\bibfnamefont {L.}~\bibnamefont {Lugiato}},\ }\href@noop {} {\bibfield
  {journal} {\bibinfo  {journal} {Phys. Rev. Lett.}\ }\textbf {\bibinfo
  {volume} {94}} (\bibinfo {year} {2005})}\BibitemShut {NoStop}%
\bibitem [{\citenamefont {Chen}\ \emph {et~al.}(2010)\citenamefont {Chen},
  \citenamefont {Agafonov}, \citenamefont {Luo}, \citenamefont {Liu},
  \citenamefont {Xian}, \citenamefont {Chekhova},\ and\ \citenamefont
  {Wu}}]{che10}%
  \BibitemOpen
  \bibfield  {author} {\bibinfo {author} {\bibfnamefont {X.}~\bibnamefont
  {Chen}}, \bibinfo {author} {\bibfnamefont {I.~N.}\ \bibnamefont {Agafonov}},
  \bibinfo {author} {\bibfnamefont {K.}~\bibnamefont {Luo}}, \bibinfo {author}
  {\bibfnamefont {Q.}~\bibnamefont {Liu}}, \bibinfo {author} {\bibfnamefont
  {R.}~\bibnamefont {Xian}}, \bibinfo {author} {\bibfnamefont {M.~V.}\
  \bibnamefont {Chekhova}}, \ and\ \bibinfo {author} {\bibfnamefont
  {L.}~\bibnamefont {Wu}},\ }\href@noop {} {\bibfield  {journal} {\bibinfo
  {journal} {Opt. Lett.}\ }\textbf {\bibinfo {volume} {35}},\ \bibinfo {pages}
  {1166} (\bibinfo {year} {2010})}\BibitemShut {NoStop}%
\bibitem [{\citenamefont {Simon}\ \emph {et~al.}(2014)\citenamefont {Simon},
  \citenamefont {Jaeger},\ and\ \citenamefont {Sergienko}}]{sim14}%
  \BibitemOpen
  \bibfield  {author} {\bibinfo {author} {\bibfnamefont {D.~S.}\ \bibnamefont
  {Simon}}, \bibinfo {author} {\bibfnamefont {G.}~\bibnamefont {Jaeger}}, \
  and\ \bibinfo {author} {\bibfnamefont {A.~V.}\ \bibnamefont {Sergienko}},\
  }\href@noop {} {\bibfield  {journal} {\bibinfo  {journal} {Int. Journ. of
  Quant. Inf.}\ }\textbf {\bibinfo {volume} {12}},\ \bibinfo {pages} {1430004}
  (\bibinfo {year} {2014})}\BibitemShut {NoStop}%
\bibitem [{\citenamefont {Gatti}\ \emph {et~al.}(1999)\citenamefont {Gatti},
  \citenamefont {Brambilla}, \citenamefont {Lugiato},\ and\ \citenamefont
  {Kolobov}}]{gat99}%
  \BibitemOpen
  \bibfield  {author} {\bibinfo {author} {\bibfnamefont {A.}~\bibnamefont
  {Gatti}}, \bibinfo {author} {\bibfnamefont {E.}~\bibnamefont {Brambilla}},
  \bibinfo {author} {\bibfnamefont {L.~A.}\ \bibnamefont {Lugiato}}, \ and\
  \bibinfo {author} {\bibfnamefont {M.~I.}\ \bibnamefont {Kolobov}},\
  }\href@noop {} {\bibfield  {journal} {\bibinfo  {journal} {Phys. Rev. Lett}\
  }\textbf {\bibinfo {volume} {83}},\ \bibinfo {pages} {1763} (\bibinfo {year}
  {1999})}\BibitemShut {NoStop}%
\bibitem [{\citenamefont {Tan}\ \emph {et~al.}(2008)\citenamefont {Tan},
  \citenamefont {Erkmen}, \citenamefont {Giovannetti}, \citenamefont {Guha},
  \citenamefont {Lloyd}, \citenamefont {Maccone}, \citenamefont {Pirandola},\
  and\ \citenamefont {Shapiro}}]{tan08}%
  \BibitemOpen
  \bibfield  {author} {\bibinfo {author} {\bibfnamefont {S.~H.}\ \bibnamefont
  {Tan}}, \bibinfo {author} {\bibfnamefont {B.~I.}\ \bibnamefont {Erkmen}},
  \bibinfo {author} {\bibfnamefont {V.}~\bibnamefont {Giovannetti}}, \bibinfo
  {author} {\bibfnamefont {S.}~\bibnamefont {Guha}}, \bibinfo {author}
  {\bibfnamefont {S.}~\bibnamefont {Lloyd}}, \bibinfo {author} {\bibfnamefont
  {L.}~\bibnamefont {Maccone}}, \bibinfo {author} {\bibfnamefont
  {S.}~\bibnamefont {Pirandola}}, \ and\ \bibinfo {author} {\bibfnamefont
  {J.~H.}\ \bibnamefont {Shapiro}},\ }\href@noop {} {\bibfield  {journal}
  {\bibinfo  {journal} {Phys. Rev. Lett}\ }\textbf {\bibinfo {volume} {101}},\
  \bibinfo {pages} {253601} (\bibinfo {year} {2008})}\BibitemShut {NoStop}%
\bibitem [{\citenamefont {Meda}\ \emph {et~al.}(2013)\citenamefont {Meda},
  \citenamefont {Olivares}, \citenamefont {Degiovanni}, \citenamefont {Brida},
  \citenamefont {Genovese},\ and\ \citenamefont {Paris}}]{med13}%
  \BibitemOpen
  \bibfield  {author} {\bibinfo {author} {\bibfnamefont {A.}~\bibnamefont
  {Meda}}, \bibinfo {author} {\bibfnamefont {S.}~\bibnamefont {Olivares}},
  \bibinfo {author} {\bibfnamefont {I.~P.}\ \bibnamefont {Degiovanni}},
  \bibinfo {author} {\bibfnamefont {G.}~\bibnamefont {Brida}}, \bibinfo
  {author} {\bibfnamefont {M.}~\bibnamefont {Genovese}}, \ and\ \bibinfo
  {author} {\bibfnamefont {M.~G.~A.}\ \bibnamefont {Paris}},\ }\href@noop {}
  {\bibfield  {journal} {\bibinfo  {journal} {Opt. Lett.}\ }\textbf {\bibinfo
  {volume} {38}} (\bibinfo {year} {2013})}\BibitemShut {NoStop}%
\bibitem [{\citenamefont {Glauber}(1963)}]{gla63}%
  \BibitemOpen
  \bibfield  {author} {\bibinfo {author} {\bibfnamefont {R.~J.}\ \bibnamefont
  {Glauber}},\ }\href {\doibase 10.1103/PhysRev.130.2529} {\bibfield  {journal}
  {\bibinfo  {journal} {Phys. Rev.}\ }\textbf {\bibinfo {volume} {130}},\
  \bibinfo {pages} {2529} (\bibinfo {year} {1963})}\BibitemShut {NoStop}%
\bibitem [{\citenamefont {Mandel}\ and\ \citenamefont {Wolf}(1995)}]{man95}%
  \BibitemOpen
  \bibfield  {author} {\bibinfo {author} {\bibfnamefont {L.}~\bibnamefont
  {Mandel}}\ and\ \bibinfo {author} {\bibfnamefont {E.}~\bibnamefont {Wolf}},\
  }\href@noop {} {\emph {\bibinfo {title} {Optical Coherence and Quantum
  Optics}}}\ (\bibinfo  {publisher} {Cambridge University Press},\ \bibinfo
  {year} {1995})\BibitemShut {NoStop}%
\bibitem [{\citenamefont {Erkmen}\ and\ \citenamefont {Shapiro}(2009)}]{erk09}%
  \BibitemOpen
  \bibfield  {author} {\bibinfo {author} {\bibfnamefont {B.~I.}\ \bibnamefont
  {Erkmen}}\ and\ \bibinfo {author} {\bibfnamefont {J.~H.}\ \bibnamefont
  {Shapiro}},\ }\href@noop {} {\bibfield  {journal} {\bibinfo  {journal} {Phys.
  Rev. A}\ }\textbf {\bibinfo {volume} {79}} (\bibinfo {year}
  {2009})}\BibitemShut {NoStop}%
\bibitem [{\citenamefont {Agafonov}\ \emph {et~al.}()\citenamefont {Agafonov},
  \citenamefont {Chekhova},\ and\ \citenamefont {Penin}}]{aga09}%
  \BibitemOpen
  \bibfield  {author} {\bibinfo {author} {\bibfnamefont {I.~N.}\ \bibnamefont
  {Agafonov}}, \bibinfo {author} {\bibfnamefont {M.~V.}\ \bibnamefont
  {Chekhova}}, \ and\ \bibinfo {author} {\bibfnamefont {A.~N.}\ \bibnamefont
  {Penin}},\ }\href@noop {} {\bibinfo  {journal} {arXiv:0911.3718v2}\
  }\BibitemShut {NoStop}%
\bibitem [{\citenamefont {Basano}\ and\ \citenamefont
  {Ottonello}(2007)}]{bas07}%
  \BibitemOpen
\bibfield  {journal} {  }\bibfield  {author} {\bibinfo {author} {\bibfnamefont
  {L.}~\bibnamefont {Basano}}\ and\ \bibinfo {author} {\bibfnamefont
  {P.}~\bibnamefont {Ottonello}},\ }\href@noop {} {\bibfield  {journal}
  {\bibinfo  {journal} {Opt. Express}\ }\textbf {\bibinfo {volume} {15}}
  (\bibinfo {year} {2007})}\BibitemShut {NoStop}%
\bibitem [{\citenamefont {Arecchi}(1965)}]{are65}%
  \BibitemOpen
  \bibfield  {author} {\bibinfo {author} {\bibfnamefont {F.~T.}\ \bibnamefont
  {Arecchi}},\ }\href {http://link.aps.org/doi/10.1103/PhysRevLett.15.912}
  {\bibfield  {journal} {\bibinfo  {journal} {Phys. Rev. Lett.}\ }\textbf
  {\bibinfo {volume} {15}},\ \bibinfo {pages} {912} (\bibinfo {year}
  {1965})}\BibitemShut {NoStop}%
\bibitem [{\citenamefont {Brida}\ \emph
  {et~al.}(2009{\natexlab{a}})\citenamefont {Brida}, \citenamefont {Caspani},
  \citenamefont {Gatti}, \citenamefont {Genovese}, \citenamefont {Meda},\ and\
  \citenamefont {Ruo-Berchera}}]{bri09}%
  \BibitemOpen
  \bibfield  {author} {\bibinfo {author} {\bibfnamefont {G.}~\bibnamefont
  {Brida}}, \bibinfo {author} {\bibfnamefont {L.}~\bibnamefont {Caspani}},
  \bibinfo {author} {\bibfnamefont {A.}~\bibnamefont {Gatti}}, \bibinfo
  {author} {\bibfnamefont {M.}~\bibnamefont {Genovese}}, \bibinfo {author}
  {\bibfnamefont {A.}~\bibnamefont {Meda}}, \ and\ \bibinfo {author}
  {\bibfnamefont {I.}~\bibnamefont {Ruo-Berchera}},\ }\href@noop {} {\bibfield
  {journal} {\bibinfo  {journal} {Phys. Rev. Lett}\ }\textbf {\bibinfo {volume}
  {102}} (\bibinfo {year} {2009}{\natexlab{a}})}\BibitemShut {NoStop}%
\bibitem [{\citenamefont {Magni}\ and\ \citenamefont {Vertesy}(2000)}]{mag00}%
  \BibitemOpen
  \bibfield  {author} {\bibinfo {author} {\bibfnamefont {A.}~\bibnamefont
  {Magni}}\ and\ \bibinfo {author} {\bibfnamefont {G.}~\bibnamefont
  {Vertesy}},\ }\href@noop {} {\bibfield  {journal} {\bibinfo  {journal} {Phys.
  Rev. B}\ }\textbf {\bibinfo {volume} {61}},\ \bibinfo {pages} {3203}
  (\bibinfo {year} {2000})}\BibitemShut {NoStop}%
\bibitem [{\citenamefont {Brida}\ \emph
  {et~al.}(2009{\natexlab{b}})\citenamefont {Brida}, \citenamefont {Genovese},
  \citenamefont {Meda}, \citenamefont {Predazzi},\ and\ \citenamefont
  {Berchera}}]{bri09A}%
  \BibitemOpen
  \bibfield  {author} {\bibinfo {author} {\bibfnamefont {G.}~\bibnamefont
  {Brida}}, \bibinfo {author} {\bibfnamefont {M.}~\bibnamefont {Genovese}},
  \bibinfo {author} {\bibfnamefont {A.}~\bibnamefont {Meda}}, \bibinfo {author}
  {\bibfnamefont {E.}~\bibnamefont {Predazzi}}, \ and\ \bibinfo {author}
  {\bibfnamefont {I.~R.}\ \bibnamefont {Berchera}},\ }\href@noop {} {\bibfield
  {journal} {\bibinfo  {journal} {Journ. Mod. Opt.}\ }\textbf {\bibinfo
  {volume} {56}},\ \bibinfo {pages} {2} (\bibinfo {year}
  {2009}{\natexlab{b}})}\BibitemShut {NoStop}%
\bibitem [{\citenamefont {Klauder}\ and\ \citenamefont
  {Sudarshan}(1968)}]{kla68}%
  \BibitemOpen
  \bibfield  {author} {\bibinfo {author} {\bibfnamefont {J.~R.}\ \bibnamefont
  {Klauder}}\ and\ \bibinfo {author} {\bibfnamefont {E.~C.~G.}\ \bibnamefont
  {Sudarshan}},\ }\href@noop {} {\emph {\bibinfo {title} {Fundamentals of
  Quantum Optics}}}\ (\bibinfo  {publisher} {Dover books on Physics},\ \bibinfo
  {year} {1968})\BibitemShut {NoStop}%
\bibitem [{\citenamefont {Shapiro}(2008)}]{sha08}%
  \BibitemOpen
  \bibfield  {author} {\bibinfo {author} {\bibfnamefont {J.~H.}\ \bibnamefont
  {Shapiro}},\ }\href@noop {} {\bibfield  {journal} {\bibinfo  {journal} {Phys.
  Rev. A}\ }\textbf {\bibinfo {volume} {78}},\ \bibinfo {pages} {061802}
  (\bibinfo {year} {2008})}\BibitemShut {NoStop}%
\bibitem [{\citenamefont {Bromberg}\ \emph {et~al.}(2009)\citenamefont
  {Bromberg}, \citenamefont {Katz},\ and\ \citenamefont {Silberberg}}]{bro09}%
  \BibitemOpen
  \bibfield  {author} {\bibinfo {author} {\bibfnamefont {Y.}~\bibnamefont
  {Bromberg}}, \bibinfo {author} {\bibfnamefont {O.}~\bibnamefont {Katz}}, \
  and\ \bibinfo {author} {\bibfnamefont {Y.}~\bibnamefont {Silberberg}},\
  }\href@noop {} {\bibfield  {journal} {\bibinfo  {journal} {Phys. Rev. A}\
  }\textbf {\bibinfo {volume} {79}},\ \bibinfo {pages} {053840} (\bibinfo
  {year} {2009})}\BibitemShut {NoStop}%
\end{thebibliography}%

\end{document}